\documentclass[12pt]{book}

\usepackage{times}
\usepackage[dvips]{graphicx,color}
\usepackage{makeidx,view}

\begin{document}

\BookTitle{\itshape The Universe Viewed in Gamma-Rays}
\CopyRight{\copyright 2002 by Universal Academy Press, Inc.}
\pagenumbering{arabic}

\chapter{
Multi-Wavelength Observations of 1ES1959+650 in a Flaring State}

\author{%
M. Schroedter$^{1}$ for the VERITAS Collaboration, E.Falco$^2$, O. M. Kurtanidze$^{3,4}$, and M. Nikolashvili$^3$\\%
{\it $^1$Smithsonian Astrophysical Observatory and University of Arizona, $^2$Whipple Observatory, $^3$Abastumani Observatory, $^4$ Landessternwarte Heidelberg-K\"onigstuhl and
	Astrophysikalisches Institut Potsdam}}

\section*{Abstract}

In May 2002, the BL Lacertae object (BL Lac), 1ES1959+650, was 
detected in the Very High Energy (VHE) regime by the VERITAS collaboration using
the Whipple $\gamma$-ray observatory.  During the following two months, the VERITAS collaboration observed episodes of flaring; at times, the nightly average VHE signal was three times as bright as the Crab Nebula, the standard candle in VHE astronomy. 1ES1959+650 was also monitored in the optical and X-ray bands. The X-ray light curve showed periodic behavior and correlation with the VHE and the optical bands.

\section{Introduction}

The BL Lac object, 1ES1959+650,  redshift  0.047, is an active galaxy powered by a super massive black hole. It was discovered in 1993 by comparing the X-ray/radio/optical fluxes of objects in the {\it Einstein} IPC Slew Survey (Schachter et. al. 1993).  
1ES1959+650 was first detected at VHE energies by the Utah Seven Telescope Array collaboration (Nishiyama et. al. 1999) in mid-1998. It was seen in a flaring state in the VHE  energy regime, this Spring for the first time by the Whipple Observatory (Dowdall et. al. 2002).  Following the initial detection, the object was intensely monitored by the VERITAS collaboration during the next two months (Holder et. al. 2002a and Holder et. al. 2002b). Overlapping with the VHE observations were R-band photometry, optical spectroscopy, and X-ray observations. The study of the cross-correlation of these energy bands  is an important tool in restricting the possible emission and absorption processes of VHE photons.
\begin{figure}[t]
  \begin{center}
    \includegraphics[height=23pc,clip]{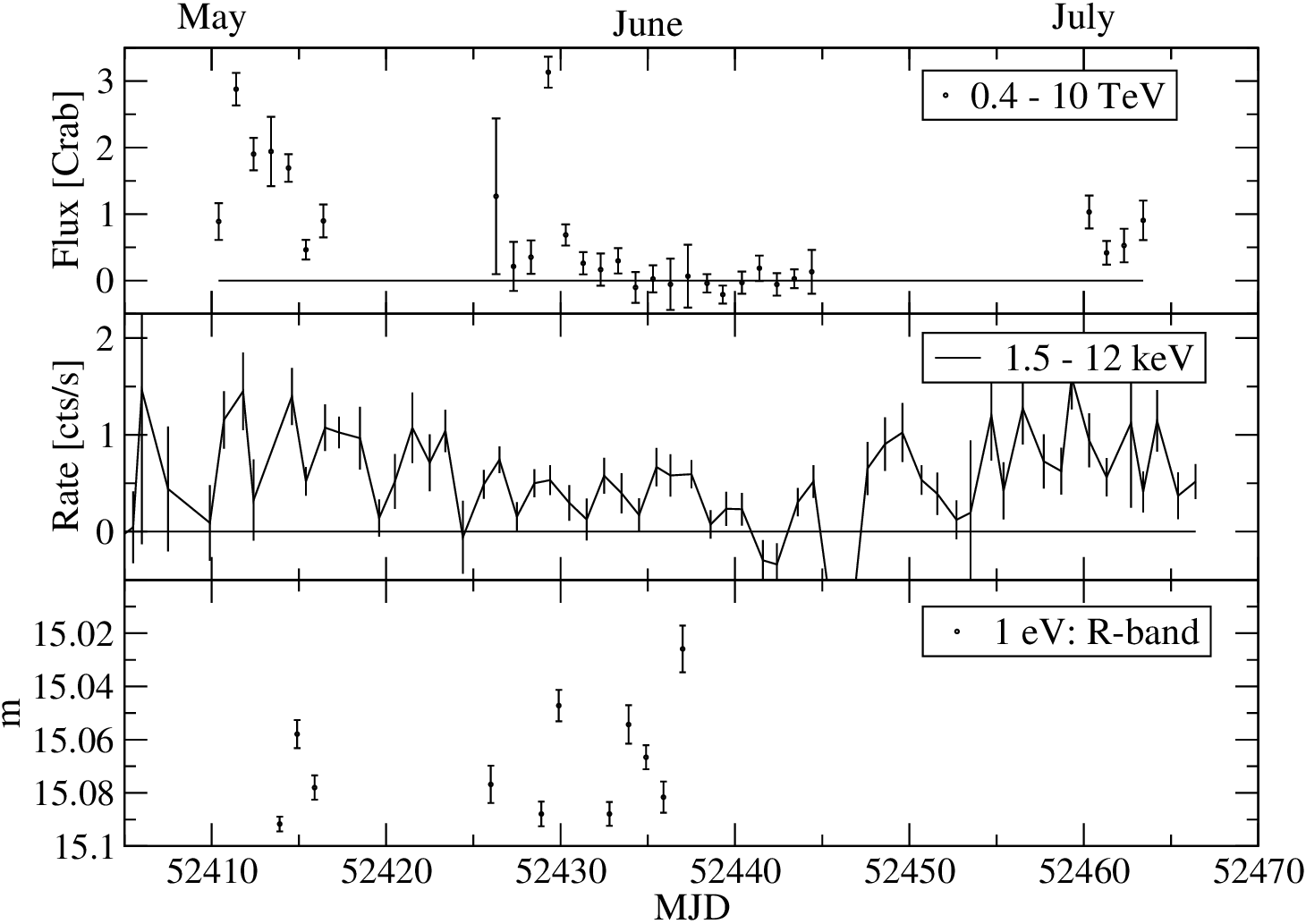}
  \end{center}
  \caption{Light curve of 1ES1959+650 in the VHE, X-ray, and optical energy bands averaged over 24 hours.}
\end{figure}

\section{Observations and Data Reduction}
\subsection{VHE observations}
The Whipple Observatory 10 m reflector images Cerenkov light flashes from air showers initiated by VHE photons in the atmosphere. The telescope is located on Mt. Hopkins near Tucson, Arizona. It consists of 238 mirror facets arranged on a spherical support structure of Davis-Cotton design, with focal length 7.3 m. The camera consists of 490 ultraviolet-sensitive photomultiplier tubes, giving a \mbox{3.8 $^\circ$} field of view. Only the inner PMTs covering \mbox{2.6 $^\circ$} are used in this analysis.
Observations on 1ES1959+650 were taken from 16 May 2002  through 8 July 2002 (UT) for a total of 36 hrs. ON source and 6.4 hrs. OFF source. The detection significance for ON/OFF paired observations is 8 $\sigma$ with an average photon rate of $1.5\pm0.2$ $\gamma$/min. Standard analysis of the data was done by moment fitting (Hillas 1985) of the events followed by a set of cuts optimized on Crab Nebula observations to optimize the signal-to-noise ratio. The light curve Fig[1] was calculated from ON source data, i.e. tracking analysis, with background estimated from OFF source data. The rate was then corrected empirically for elevation and throughput (LeBohec and Holder 2002) based on observations of the Crab Nebula, the standard candle at VHE energies, taken from February to April 2002.  Strictly, this is only valid if the sources have the same spectral index $\gamma$; $\gamma_{Crab}=2.5$. 
\begin{figure}[t]
  \begin{center}
    \includegraphics[height=20pc,clip]{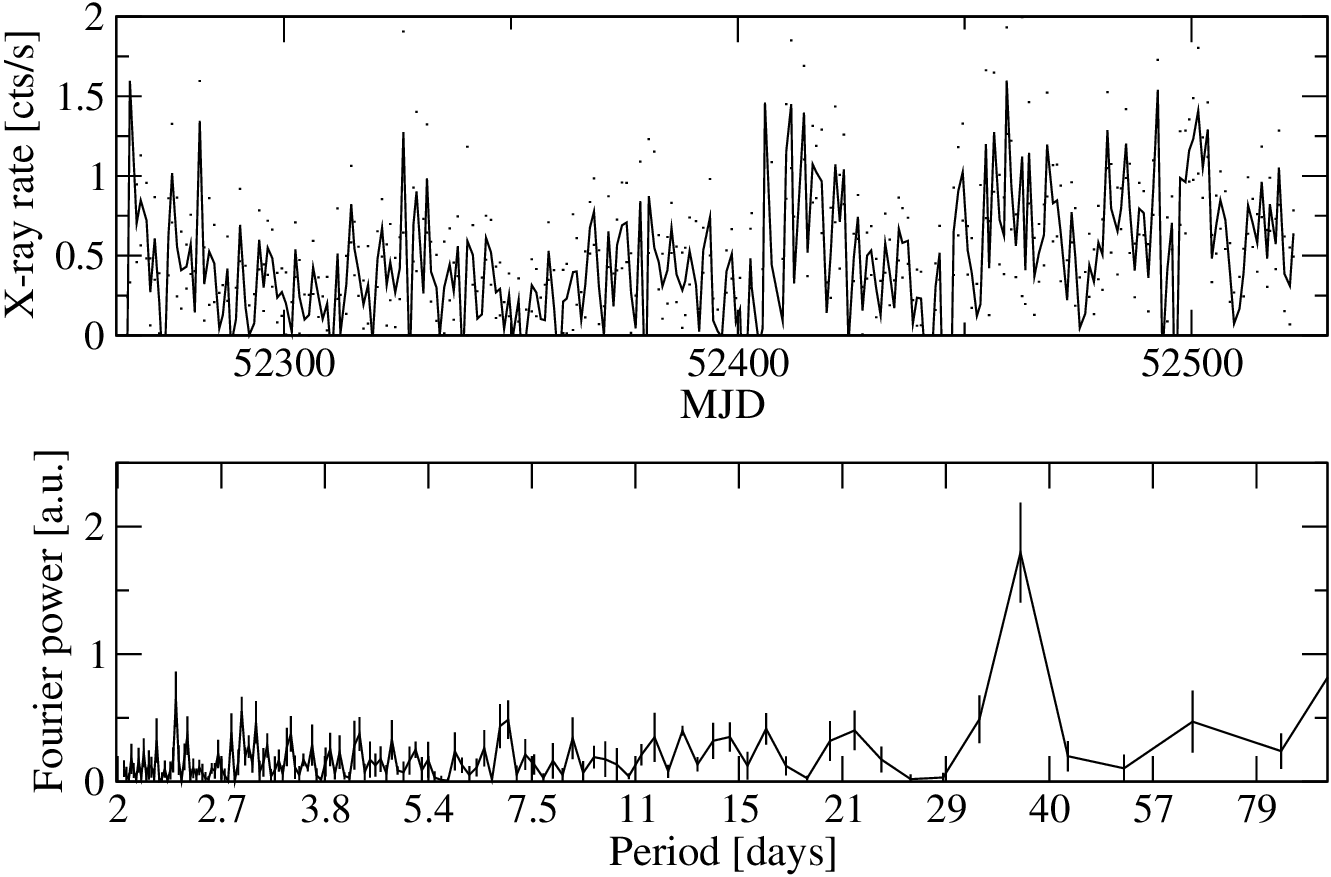}
  \end{center}
  \caption{Top: ASM X-ray light curve, 1 $\sigma$ errors indicated by dots. Bottom: Fourier transform of light curve.}
\end{figure}

\subsection{ X-ray observations}
X-ray measurements were taken by the ASM detector on board the RXTE satellite. The ASM quick-look results were provided by the ASM/RXTE team. They calculated the 24-hour average flux  from five to ten 90-second dwells. The ASM detects photons in  the energy range from 1.5 keV to 12 keV, with peak sensitivity at 5 keV.

\subsection{Optical observations}
The R-band (650 nm) photometry was taken with the 70 cm telescope at the Abastumani Observatory in Georgia from 19 May to 11 June 2002. The frames were reduced using DAOPHOT II. The apparent magnitude of 1ES1959+650 was determined by comparison with 3 standard stars in the field of view (Villata et. al.1998).
Optical spectra were taken using the newly refurbished 6.5 m telescope at the MMT. The spectra were taken while 1ES1959+650 was active at VHE energies and one week later when the flux was not detectable. The first of the two spectra was measured on 5 June 2002 with a 300 lines/mm grating and 45 min exposure, the second spectrum was taken on 13 June 2002 with a 500 lines/mm grating and 10 min exposure. Not all cosmic ray contaminants were removed from the spectra.

\section{Light curve correlations}

 \subsection{Quasi-periodic X-ray flaring  }
The X-ray light curve, Fig [2], shows a series of flares spaced roughly 40 days apart. Also shown is the associated Fourier transform, Fig[2]. It shows that there is weak evidence for a ~40 day period of X-ray flaring. A similar pattern of flaring activity was possibly found for Mkn 501 (Kranich et. al. 1999).

\subsection{VHE and X-ray correlation}
The X-ray - VHE linear correlation coefficient (LCC) as a function of the time delay between the bands is shown in Fig[3]. 
Simultaneous keV-VHE measurements are correlated with a LCC = 0.45. Here, simultaneous means $\pm12$ hours. To determine the significance of this correlation, the keV and VHE flux measurements were randomized 100 times with respect to one another and the LCC was calculated for each random set. This resulted in a mean LCC = 0.1 $\pm$ 0.1, so we deduce that the correlation between X-ray and VHE flux is significant at the level of 3.5 $\sigma$. Assuming the LCC distribution to be Gaussian, the probability of at least one 3.5 $\sigma$ deviation in 61 independent samples is 0.03. This is an approximation as neither the LCC distribution is Gaussian nor the 61 time intervals are completely independent samples.
On the other hand, when the X-ray flux is time delayed by 25 days following the VHE flux, we find an anti-correlation of the X-ray and VHE flux. The LCC factor is 0.42, the probability for this to happen by chance is 0.15. 
So, there is weak evidence for both a 40 day periodicity of the X-ray flux and simultaneous correlation with VHE photon emission around the time of observation.  A simultaneous correlation of X-ray and VHE flux has also been found during flares of Mrk 421 in 1995 and Mrk 501 in 1997 (Buckley 1999).
\begin{figure}[t]
  \begin{center}
    \includegraphics[height=23pc,clip]{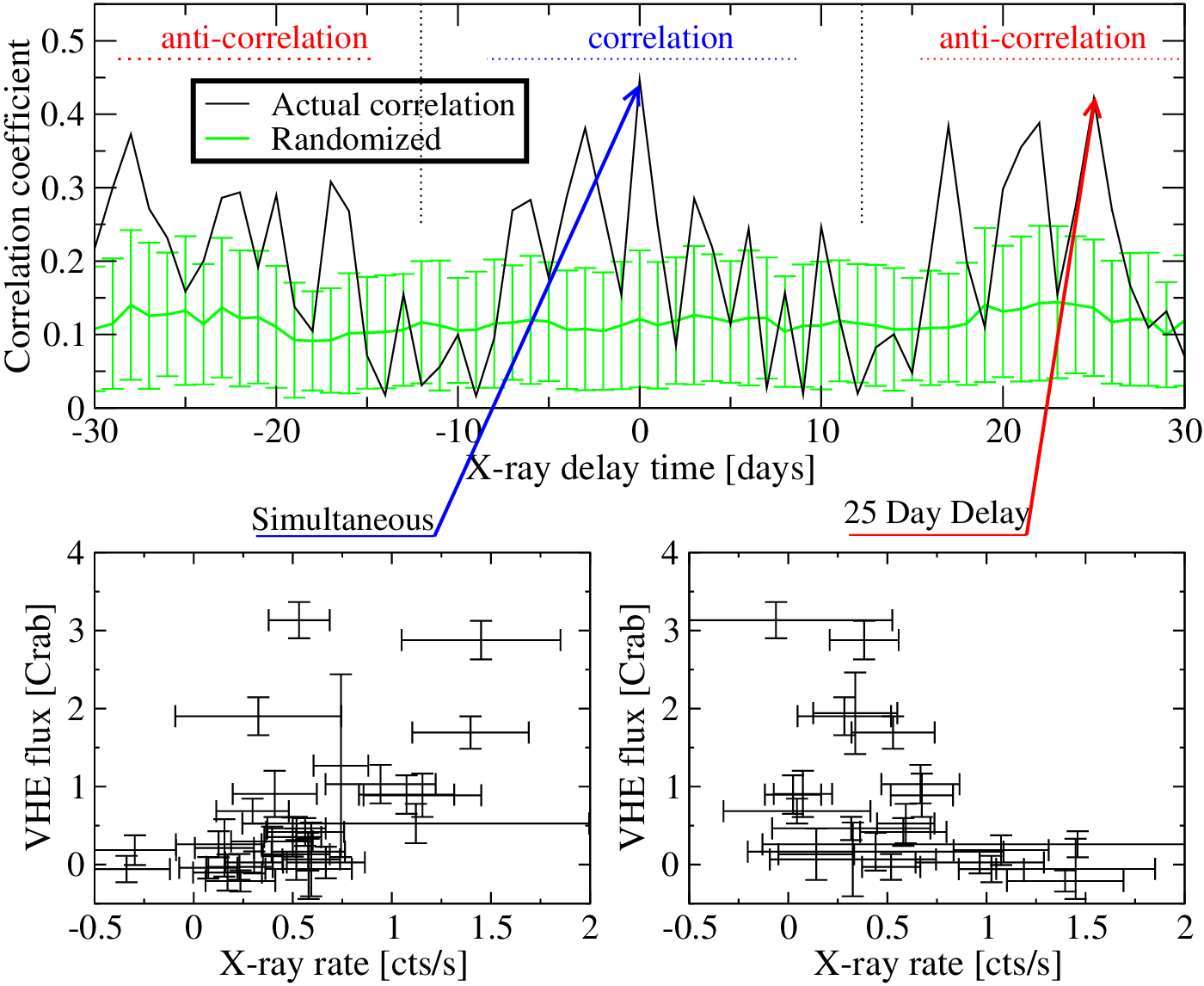}
  \end{center}
  \caption{Correlation of X-ray and VHE rate.}
\end{figure}

\subsection{Optical and X-ray correlation }
In the optical, 1ES1959+650 was very dim compared to the light curves from 1996-97 (Villata et. al. 2000) and 1997-99 (Kurtanidze et. al. 1999). During those times the magnitude varied from 15.05 to 14.55. Thus, during our VHE detection the source was optically fainter and not as variable as during previous observations. 
The X-ray and optical (R-band) emission appear to be correlated, although with the latter delayed or lagging by 5 days. The LCC is 0.78, 5 $\sigma$ away from a Gaussian random distribution. The probability for this to occur randomly in 16 independent samples is $9\times10^{-6}$. Delayed correlation between R-band and X-ray emission has also been seen in a flare of Mrk421 in 1996 (Buckley 1999).
\begin{figure}[t]
  \begin{center}
    \includegraphics[height=12pc,clip]{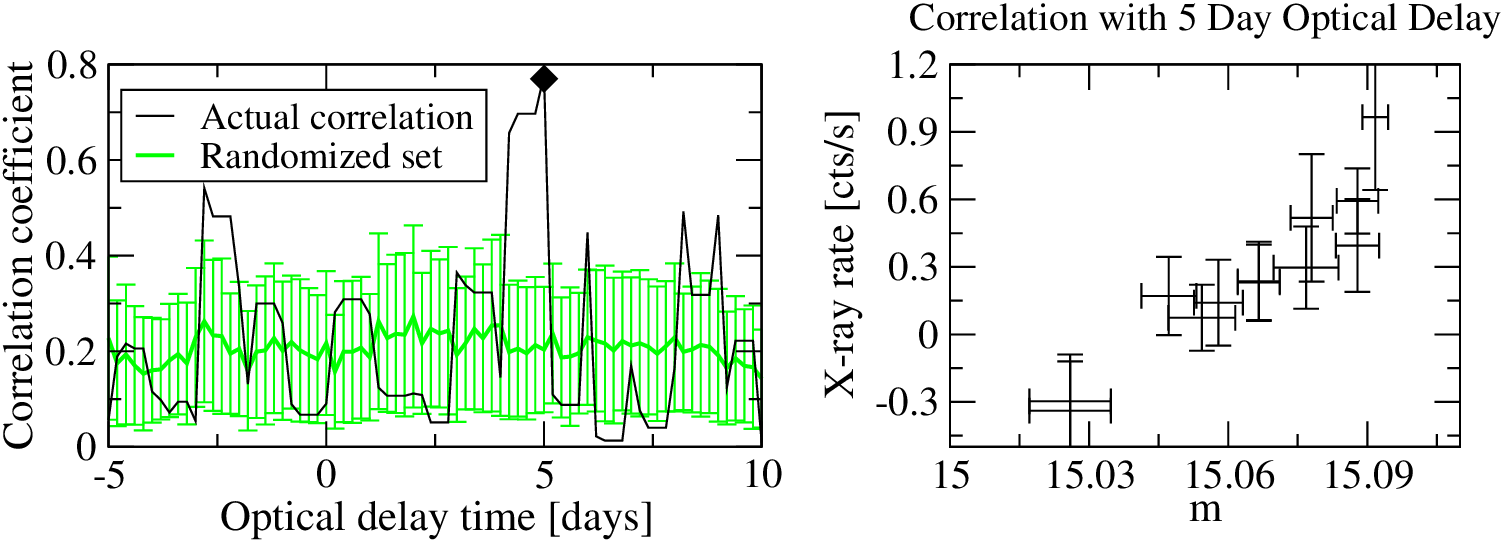}
  \end{center}
  \caption{Left: The dependence of the  R-band - VHE linear correlation coefficient on the time delay between the bands. Right: Correlation with a 5-day delayed optical magnitude.}
\end{figure}


\section{ Optical spectra}
Typical of a BL Lac object, the optical spectrum of 1ES1959+650 was nearly featureless. The spectra were neither flux calibrated nor corrected for atmospheric extinction.
The relative change in spectral features was calculated by first normalizing to the smoothed continuum and then subtracting the two normalized spectra from one another. The smoothed continuum was calculated by first applying a 3 pixel wide boxcar average to the 300 lines/mm spectrum and a 5 pixel average to the 500 lines/mm spectrum to minimize any effects the different gratings had. Then, the continuum was calculated as a 100 pixel (200\AA) boxcar average. 
The relative change in the optical spectrum between states of high VHE emission and low VHE emission was measured after applying a 3-pixel box smoothing to reduce the sensitivity to noise and possible mismatched wavelength calibration. The intensity of the following lines showed a decrease from 5 June to 13 June: $H_\alpha$ (gfwhm = 9 \AA) 3\%, $Mg$ (gfwhm = 9 \AA) 2.5\% , and $CaFe$ (gfwhm = 11 \AA) 2 \%. Unfortunately, the precision of the measurement for the $H_\alpha$ is uncertain because it is located in a region of strong night-sky absorption lines.

 \section{Summary}
The BL Lac object, 1ES1959+650, was detected in Spring 2002 with the Whipple 10m Cherenkov Imaging Telescope. The 24-hour average VHE photon flux was measured to vary by as much as 3 Crab units.  The VHE and X-ray flux showed a slight simultaneous correlation. There is evidence for a roughly 40 day period in the X-ray light curve. A strong correlation between the 5-day delayed optical R-band brightness and the X-ray flux was also observed. The R-band brightness was fainter than during previous observations and optical spectra were without any prominent features or variability.

\section*{References}

\vspace{1pc}
\noindent
1.\ Buckley J.H. 1999, ApJ, 11, 119
\re
2.\ Dowdall C., Moriarty P., and Kosack K. 2002, IAU Circular 7903
\re
3.\ Hillas A.M. 1985, Proc. 19th ICR, 3, 445
\re
4.\ Holder J. et. al. 2002a, these proceedings, Kashiwa, Japan 2002
\re
5.\ Holder J. et. al. 2002b, astro-ph/0212170,  and ApJ in press 
\re
6.\ Kranich D., et. al, 1999, 26th ICRC, 3, 358
\re
7.\ Kurtanidze O. M., et. al. 1999, OJ-94 Ann. meeting, Torino, Italy, 29-32
\re
8.\ LeBohec S. and Holder J. 2002, ApJ, in press
\re
9.\ Nishiyama T., et. al. 1999, 26th ICRC, 3, 370
\re
10.\ Schachter J.F. et. al. 1993, ApJ, 412, 541
\re
11.\ Villata M., et. al. 1998, A\&AS 130, 305
\re
12.\ Villata M., et. al. 2000, A\&AS144, 481




%

\endofpaper

\end{document}